\documentclass[12pt]{article} 

\usepackage{xcolor}

\pdfoutput=1
\usepackage{amssymb,graphicx}
\usepackage{epstopdf}
\usepackage{amsmath,amsfonts}
\usepackage{epsfig} 
\usepackage{graphicx,graphics}

\renewcommand{\l}{\left(}
\renewcommand{\r}{\right)}

\begin{document}

\sloppy

\title{\bf New mechanism of producing superheavy Dark Matter}
\author{Eugeny Babichev$^{a}$, Dmitry Gorbunov$^{b,c}$, and Sabir Ramazanov$^{d}$\\
 \small{$^a$\em Laboratoire de Physique Th\'eorique, CNRS,} \\ 
 \small{\em  Univ. Paris-Sud, Universit\'e Paris-Saclay, 91405 Orsay, France}\\
  \small{$^b$\em Institute for Nuclear Research of the Russian Academy of Sciences,}\\
\small{\em 60th October Anniversary prospect 7a, Moscow 117312, Russia}\\
\small{$^c$ \em Moscow Institute of Physics and Technology, Institutsky per. 9, Dolgoprudny 141700, Russia}\\
\small{$^d$ \em CEICO, Institute of Physics of the Czech Academy of Sciences,}\\
\small{\em Na Slovance 1999/2, 182 21 Prague 8, Czech Republic}
 }

{\let\newpage\relax\maketitle}

\begin{abstract}
We study in detail the recently proposed mechanism of generating
superheavy Dark Matter with the mass larger than the Hubble rate at
the end of inflation. A real scalar field constituting Dark Matter
linearly couples to the inflaton. As a result of this
  interaction, the scalar gets displaced from its zero expectation
  value. This offset feeds into the energy density of Dark
  Matter. This mechanism is universal and can be implemented in a
  generic inflationary scenario. Phenomenology of the model is
comprised of Dark Matter 
decay into inflatons, which in turn decay into Standard Model
species triggering cascades of high energy particles contributing to
the cosmic ray flux. We evaluate the lifetime of Dark Matter and obtain limits on the inflationary scenarios, where this mechanism does not lead to the conflict with the 
Dark Matter stability considerations/studies of cosmic ray propagation. 

\end{abstract}

\section{Introduction} 

There is a large variety of Dark Matter (DM) candidates with masses
spanning many orders of magnitude. So far, experimental and
observational searches mainly focused on the electroweak scale
DM. However, non-observation of deviations from the Standard Model of
particle physics (SM) at those scales motivates looking for more
``exotic'' candidates. In the present work, we discuss superheavy DM
with the masses larger than the Hubble rate during inflation.

Superheavy DM can be created gravitationally at the end of
inflation~\cite{Chung:1998zb, Kuzmin:1998uv,
  Kuzmin:1998kk}\footnote{General formalism of particle creation by
  non-stationary gravitational fields was developed in
  Refs.~\cite{Grib:1970xx, Parker:1969au, Zeldovich:1971mw,
    Mamaev:1976zb}.}, during preheating~\cite{Chung:1998ua,
  Greene:1997ge} and reheating~\cite{Kuzmin:1997jua, Chung:1998rq,
  Gorbunov:2010bn}, and from the collisions of vacuum bubbles at phase
transitions~\cite{Chung:1998ua}; see Ref.~\cite{Kuzmin:1999zk} for a
review. Observational consequences of superheavy DM have been
elaborated in Refs.~\cite{Berezinsky:1997hy, Berezinsky:1998ft}. Here
we discuss another production mechanism, where DM modeled by a real
scalar field $\phi$ is generated through a linear coupling to some
function of an inflaton~\cite{Babichev:2018afx, Babichev:2018sia}. In
that case, the field $\phi$ acquires an effective non-zero expectation
value during inflation (Section~2). After inflation, this expectation
value sets the amplitude of the field $\phi$ oscillations. From that
moment on, the evolution of the $\phi$-condensate averaged over many
oscillations is that of the pressureless fluid serving as DM. 
  This is in spirit a particular realization of the vacuum
  misalignment mechanism underlying axionic
  models~\cite{Marsh:2015xka} or string inflationary scenarios
  involving moduli fields~\cite{Cicoli:2016olq}. Now we apply the
  mechanism to superheavy fields.

This mechanism of superheavy particles generation
\cite{Babichev:2018afx} is different from the known ones in many
aspects. First, in our scenario non-zero energy density of DM is
already present at the stage of inflationary expansion of the
Universe\footnote{This is different compared to the resonant
  production of particles during inflation discussed in
  Ref.~\cite{Chung:1999ve}. There the concentration of created
  particles gets diluted by the inflationary expansion, unless the
  production takes place at the last e-foldings of inflation. In our
  case, the concentration of particles once produced is kept constant
  all the way down to the end of inflation.}. Second, with our
  mechanism there is no a priori upper bound on the mass of produced
  particles (still it should be below the Planck
  scale). This is in contrast to gravitational
production, where the masses of superheavy particles are fixed by the
Universe expansion law \cite{Mamaev:1976zb, Ema:2018ucl} or equal a
few times the Hubble rate at the end of inflation~\cite{Chung:1998zb,
  Kuzmin:1998uv, Kuzmin:1998kk, Chung:2004nh}, and scenarios of DM
creation at reheating, where possible masses are limited by the
reheating temperature~\cite{Chung:1998rq}. Third, as it follows from
the above discussion, our mechanism does not require a thermal bath,
where the DM particles would be produced through the scattering
processes.  Our scenario only requires the existence of the inflaton
condensate feebly coupled to $\phi$, while the scattering cross
section of $\phi$-particles can be negligible.

Linear interaction of the superheavy field $\phi$ with the inflaton
implies that it is generically unstable. We calculate the lifetime of
DM (Section~3). The simplest case with the renormalizable interaction
between the field $\phi$ and the inflaton taking Planckian values, and
the largest possible expansion rate (suggesting relic gravitational
waves detectable in the future experiments) is only marginally
consistent with the DM lifetime to be of the order of the present age
of the Universe. For non-renormalizable interactions and/or lower
scale inflation, the stability constraint is satisfied in some range
of DM masses. Generically, however, the stability is not a sufficient
condition, because the inflatons decay into SM species (this coupling
is needed to reheat the post-inflationary Universe) producing cascades
of high energy particles. In particular, studies of the
gamma-rays~\cite{Ackermann:2014usa} and IceCube
neutrinos~\cite{Aartsen:2014gkd} set more severe limits on the DM
lifetime, which should exceed the age of the Universe by many,
typically $10-12$, orders of magnitude. See
Refs.~\cite{Esmaili:2012us, Aloisio:2015lva,Kalashev:2016cre, Marzola:2016hyt,
  Cohen:2016uyg, Kachelriess:2018rty, Blanco:2018esa} for the state of
the art and Ref.~\cite{Ibarra:2013cra} for the review. Thus, within
the new suggested mechanism of DM production there is an interesting
opportunity to relate the properties of the inflationary models with
the observations of high energy cosmic rays.

\section{The model}
We are interested in the model with the Lagrangian
\begin{equation}
\label{actiongeneral}
{\cal L}=\frac{1}{2}g^{\mu\nu}\partial_\mu\phi\partial_\nu\phi-\frac{1}{2}M^2\phi^2+ \phi \cdot F(\varphi) \; .
\end{equation}
Here $M$ is the mass of the scalar $\phi$ constituting DM; $\varphi$ is the inflaton, which we assume to be 
a canonical scalar field with an almost flat potential; $F(\varphi)$ is some
generic function of the inflaton, such that $F(\varphi) \rightarrow 0$ as $\varphi \rightarrow 0$. The background equation of motion for the field $\phi$ is that of the damped oscillator with the external force $F(\varphi)$ applied: 
\begin{equation}
\label{osc}
\ddot{\phi}+3H \dot{\phi}+M^2 \phi =F(\varphi) \; .
\end{equation}
The key idea is to consider very large masses $M$, so that 
\begin{equation}
\label{basic}
M \gtrsim H_e \; ,
\end{equation}
where $H_e$ is the Hubble rate during the last e-foldings of inflation. Once the condition~\eqref{basic} is fulfilled, the field $\phi$ quickly relaxes to its effective minimum 
\begin{equation}
\label{tracking}
\phi=\frac{F(\varphi)}{M^2} \; .
\end{equation} 

After inflation, when the inflaton $\varphi$ drops down considerably, one finds
\begin{equation}
\label{condensate}
\phi=\frac{A \cdot F (\varphi_e)}{M^2} \left(\frac{a_e}{a} \right)^{3/2} \cos \left[M(t-t_e)+\delta_e \right] \; .
\end{equation}
That is, the field $\phi$ undergoes coherent oscillations with the
frequency $M$. Here $\delta_e$ is an irrelevant phase, while the
dimensionless coefficient $A$ is a penalty factor for non-instantaneous decoupling from the inflaton. Indeed, if
post-inflationary evolution of $F(\varphi)$ is smooth and slow, the
field $\phi$ still tends to track the inflaton, which predicts
$\phi\to 0$ as $\varphi\to 0$. Were tracking exact, we would have $A=0$. However, there are always deviations from the 
tracking solution, which feed into the parameter $A$, so that $A \neq 0$. The strength of these deviations and hence the 
value of $A$ depend on the rate of the inflaton change at the end of inflation/during post-inflationary stage. 
As it follows, the resulting coefficient $A$ crucially relies on the choice of the model, and generically may depend on $M$, inflaton time-scale(s), Hubble parameter,
reheating temperature. We
illustrate this by exact calculations of the coefficients $A$ using toy examples in Appendix. The realistic situation with the fast change of the inflaton leading to relatively large $A$,
can happen at various stages: 
right at the end of inflation, as in the hybrid inflation or
in the healthy Higgs inflation \cite{Bezrukov:2019ylq}; later at
preheating dominated by anharmonic oscillations; or even at the very
beginning of the hot stage, as in models with tachyonic
preheating.

The energy density of the oscillating condensate is conserved in the
comoving volume: 
\begin{equation}
\label{abundance}
 \rho_{\phi} =\frac{A^2 F^2(\varphi_e)}{2M^2} \cdot \left(\frac{a_e}{a} \right)^3\; .
\end{equation}
After reheating the ratio of $\phi$-particle energy density to the
entropy density $s(T)$ remains constant:
\begin{equation}
\label{entropy}
 \frac{\rho_{\phi}(T)}{s(T)} =\text{const}\, .
\end{equation}
We assume that $\phi$-particles constitute most of the invisible
matter in the late Universe, so that we have
\begin{equation}
\label{mradeq} 
\rho_{\phi} (T_{eq}) \approx \rho_{rad} (T_{eq}) \,,
\end{equation}
where the subscript $'eq'$ stands for the matter-radiation equality, which happened at plasma temperature $T_{eq}\approx 0.8$\,eV. We ignore
the subdominant contribution of baryons to the matter density. 
The thermal radiation energy density and
entropy density are given by 
\begin{equation}
\label{rad}
\rho_{rad}(T) =\frac{\pi^2 g_*(T)}{30} T^4 \; ,\;\;\;\;s(T) =\frac{2\pi^2 h_*(T)}{45} T^3 \; ,
\end{equation} 
where $g_*(T)$ and $h_*(T)$ are the corresponding effective numbers of
ultra-relativistic degrees of freedom. In the 'standard' particle cosmology they
coincide at $T\gg 1$\,MeV, but differ at low temperature because of
neutrino decoupling, $h_*(T_{eq})\approx 3.9$,  $g_*(T_{eq})\approx
3.4$; see, e.g., Ref.~\cite{Gorbunov:2011zz}. 

Using Eq.~\eqref{entropy} and plugging Eqs.~\eqref{abundance} and~\eqref{rad} into Eq.~\eqref{mradeq}, we obtain 
\begin{equation}
\label{abundance2} 
\frac{15\, A^2\, F^2_e\,h_*(T_{eq})}{\pi^2\, M^2\,g_*(T_{reh})\,
  T^3_{reh}\, g_*(T_{eq})\,T_{eq}}
\cdot \left(\frac{a_e}{a_{reh}} \right)^3
= \frac{A^2\, F^2_e\,h_*(T_{eq})}{2\,M^2\,g_*(T_{eq})}\frac{1}{\rho_{reh}}\frac{T_{reh}}{T_{eq}}
\cdot \left(\frac{a_e}{a_{reh}} \right)^3
\approx 1 \; ,
\end{equation}
where the subscript $'reh'$ stands for the reheating; $g_* (T_{reh})
\gtrsim 100$. The ratio $a_{reh}/a_{e}$ is determined by the total matter equation
of state at the stage between inflation and reheating. 
Typically, the effective pressure and energy density are proportional
to each other, $p=w \rho$. For a constant equation of state $w$, one has
\begin{equation}
\nonumber 
\frac{\rho_{e}}{\rho_{reh}}=\left(\frac{a_{reh}}{a_e} \right)^{3(1+w)} \; .
\end{equation}
Here we ignored that the energy is equally distributed 
between the inflaton and radiation at the moment of reheating. Substituting $\rho_{e} \approx \frac{3}{8\pi} H^2_e M^2_{Pl}$, we obtain 
\begin{equation}
\nonumber 
\l\frac{a_e}{a_{reh}}\r^3 \approx \l
\frac{8\pi^3 \, g_* (T_{reh}) \, T^4_{reh}}{90 \, H^2_e \, M^2_{Pl}}
\r^{\frac{1}{1+w}} \; .
\end{equation}
Hence, in the case $w=0$, the condition~\eqref{abundance2} takes the form 
\begin{equation}
\label{abundance0}
\frac{4\pi \, h_*(T_{eq})\, A^2 \, F^2_e \, T_{reh}}{3 g_*(T_{eq})\, M^2 \, H^2_e \, M^2_{Pl} \, T_{eq}} \approx 1 \; .
\end{equation}
The Planck mass is defined by $M_{Pl} \equiv G^{-1/2}_N \approx 1.22 \cdot 10^{19}~\mbox{GeV}$, where $G_N$ is the gravitational constant. If the equation of state right after inflation mimics that of
radiation, $w=1/3$, one has
\begin{equation}
\label{abundancerad}
\l\frac{320\pi}{9}\r^{1/4}\frac{ h_*(T_{eq})\,A^2 \, F^2_e}{g_*(T_{eq})\,g^{1/4}_{*} (T_{reh}) \, M^2 \, H^{3/2}_e \, M^{3/2}_{Pl} \, T_{eq}} \approx 1 \; .
\end{equation}
In this case the reheating temperature drops out of the abundance
constraint~\eqref{abundancerad}. 

For a particular inflationary scenario and a form of the function $F(\varphi)$, one can use Eq.~\eqref{abundance} to infer the strength of the coupling between the field $\phi$ and the inflaton. Knowing the coupling, one 
calculates the decay rate of DM particles into the inflatons triggered by the interaction term in the action~\eqref{actiongeneral}. 
We will see that this can be large enough to rule out or strongly constrain the model, for some well motivated inflationary models. 

The condition~\eqref{basic} generally guarantees that no $\phi$-particles are created gravitationally at the end of inflation. Indeed,  for $M \gtrsim H_e$, the number of gravitationally produced particles
is suppressed exponentially~\cite{Kuzmin:1998uv, Chung:2004nh} (see, however, Ref.~\cite{Gorbunov:2010bn}). Precise form of the suppression depends on the choice of the inflationary scenario. 
For example, in quadratic inflation with the inflaton mass $m$, the observed abundance of DM is
reached for $M \simeq 3-5 m$~\cite{Kuzmin:1998uv} for the range of reheating temperatures
$T_{reh}=10^{9}-10^{15}~\mbox{GeV}$. 
Based on this example, where $m \simeq H_e$, we assume the minimal value $M \simeq 5 H_e$ in what follows. 

Note also that with the condition~\eqref{basic} applied, isocurvature perturbations $\delta \phi_{iso}$ of the field $\phi$ are automatically suppressed. Indeed, fluctuations $\delta \phi_{iso}$ behave as a free field. Once 
the equality $M \simeq H$ is reached at some point during inflation, they start to decay as $1/a^{3/2}$. According to the estimate given above, this equality is reached at $H \simeq 5 H_e$ for the minimal possible value of $M$. This typically corresponds to tens of e-foldings 
before the end of inflation. In the example of quadratic inflation one
has $H_{60e}/H_e \simeq 6$, where $H_{60e}$ is the Hubble rate, when
presently measured cosmological modes cross the horizon, i.e., at $60$ e-foldings\footnote{The Hubble rate $H_{60e}$ is inferred from the value of the tensor-to-scalar ratio, which reads in quadratic inflation $r\approx 0.13$. Hence, $H_{60e} \approx 9 \cdot 10^{13}~\mbox{GeV}$. 
The inflaton mass is $m \approx 1.4 \cdot 10^{13}~\mbox{GeV}$. Given that $m \simeq H_e$, one gets the estimate from the text.}. Given fast expansion of the Universe at those times, isocurvature perturbations get diluted with an exponential accuracy.

\section{Decay rate into inflatons}
In this Section, we estimate the decay rate of $\phi$-particles into inflatons, provided that the abundance constraint~\eqref{abundance0} for $w=0$ (or~\eqref{abundancerad} for $w=1/3$) is fulfilled. 
The decay rate of some particle $\phi$ with the mass $M$ into $n$ indistinguishable particles is given by 
\begin{equation}
\nonumber 
\Gamma_{\phi}=\frac{1}{2M n!}\int  |{\cal M}|^2 d \Phi_n \; .
\end{equation}
Here ${\cal M}$ is the matrix element, which describes the decay, and
$d \Phi_n$ is the phase-space element: 
\begin{equation}
\nonumber 
d\Phi_n =(2\pi)^4 \cdot \delta^{(4)} \left({\cal P}-\sum_i p_i \right) \cdot \Pi^{n}_{i=1} \frac{d^3 {\bf p}_i}{2(2\pi)^3 p^{0}_i} \; ;
\end{equation}
${\cal P}$ and $p_i$ are the 4-momenta of the particle $\phi$ and the decay products, respectively. 
For the constant ${\cal M}$, the full decay rate reads
\begin{equation}
\label{decayfull}
\Gamma_{\phi}=\frac{|{\cal M}|^2}{2M n!} \cdot \int d \Phi_n \; .
\end{equation}
In the case of massless particles in the final state, the integral over the phase space is given by 
\begin{equation}
\label{phasespace}
\int d \Phi_n =\frac{1}{2 (4\pi)^{2n-3}} \cdot \frac{M^{2n-4}}{(n-1)! (n-2)!} \; .
\end{equation}
Combining Eqs.~\eqref{decayfull} and~\eqref{phasespace}, we obtain 
\begin{equation}
\label{decaygeneral}
\Gamma_{\phi}=\frac{\left|{\cal M} \right|^2 \cdot M^{2n-5}}{4 (4\pi)^{2n-3} n!\,(n-1)! \,(n-2)!} \; .
\end{equation}

We consider the powerlaw interaction in the inflaton field $\varphi$: 
\begin{equation}
\label{renorm}
F(\varphi)=\alpha \frac{\varphi^n}{\Lambda^{n-3}} \; ,
\end{equation}
where $\alpha$ is a dimensionless constant and $\Lambda$ is the parameter of the mass dimension related to the scale of new physics, e.g., the Planck mass in the case of quantum gravity or Grand Unification scale. 
The cases $n>3$, $n=3$, and $n<3$ correspond to the non-renormalizable, renormalizable and super-renormalizable interactions, respectively. 
In what follows, we focus on the former two cases. Using Eq.~\eqref{decaygeneral}, where we substitute ${\cal M}=-n! i \alpha/\Lambda^{n-3}$, we get:  
\begin{equation}
\label{nonrenorm}
\Gamma_{\phi}=\frac{ \alpha^2 \cdot n}{4 \cdot (4\pi)^{2n-3} (n-2)! } \cdot \frac{M^{2n-5}}{\Lambda^{2n-6}} \; .
\end{equation}
The coupling constant $\alpha/\Lambda^{n-3}$ is constrained by the condition~\eqref{abundance0} or~\eqref{abundancerad} depending on the 
cosmological evolution between inflation and reheating. Extracting the coupling
constant $\alpha/\Lambda^{n-3}$ from Eqs.~\eqref{abundance0}
and~\eqref{abundancerad}, and substituting it into Eq.~\eqref{nonrenorm}, we get 
for the ratio of the DM lifetime $\tau_{\phi} \equiv
\Gamma^{-1}_ {\phi}$ to the present age of the Universe $\tau_{U} \approx 1.38 \cdot 10^{10}~\mbox{years}$:
\begin{equation}
  \label{rate-0}
  \frac{\tau_{\phi}}{\tau_{U}} \approx \frac{(n-2)! }{n} \cdot 10^{12n-48} \cdot (4.9\pi)^{2n} \cdot \left(\frac{MA}{H_e} \right)^2 \cdot \left(\frac{\varphi_e}{M_{Pl}} \right)^{2n} \cdot \left(\frac{10^{13}~\mbox{GeV}}{M} \right)^{2n-1} 
\cdot \frac{T_{reh}}{10^{12}~\mbox{GeV}} 
\end{equation}
for $w =0$ and 
\begin{equation}
  \label{rate-1/3}
  \frac{\tau_{\phi}}{\tau_{U}} \approx \frac{ (n-2)! }{g^{1/4}_* (T_{reh}) \cdot n} \cdot 10^{12n-44} \cdot (4.9\pi)^{2n} \cdot \left(\frac{MA}{H_e} \right)^2 \cdot \sqrt{\frac{H_e}{M}} \cdot \left(\frac{\varphi_e}{M_{Pl}} \right)^{2n} 
\cdot \left(\frac{10^{13}~\mbox{GeV}}{M} \right)^{2n-3/2}   
\end{equation}
for $w=1/3$.

To proceed, we need to specify the coefficient $A$. As is
  emphasized in Section~2, it is strongly model-dependent and relies
  on the conditions at the end of inflation/beginning of
  post-inflationary stage. In particular it may happen that $A$ is exponentially decreasing function of $M$. Then, the lifetime $\tau_{\phi}$ is too small being in conflict with DM stability,
  unless the mass $M$ is tuned to be close to $H_e$. To illustrate the
  mechanism we are primarily
  interested in the situation, when $A$ has a powerlaw dependence on
  $M$.  For instance, this happens in the situation, when the inflaton
  $\varphi$ (and hence the function $F(\varphi)$) is constant at times
  $t<t_e$, and then decays as a powerlaw $\varphi \propto
  \left(\frac{t_e}{t} \right)^{s}$ at times $t>t_e$. Independently of
  the power $s$, which is fixed by the choice of the inflationary
  scenario, we have obtained in Ref.~\cite{Babichev:2018sia}:
\begin{equation}
\label{noninst}
A \simeq \frac{H_e}{M} \; .
\end{equation}
For convenience we re-derive this estimate in Appendix. 

\begin{figure}[tb!]
\begin{center}
\includegraphics[width=0.45\columnwidth,angle=0]{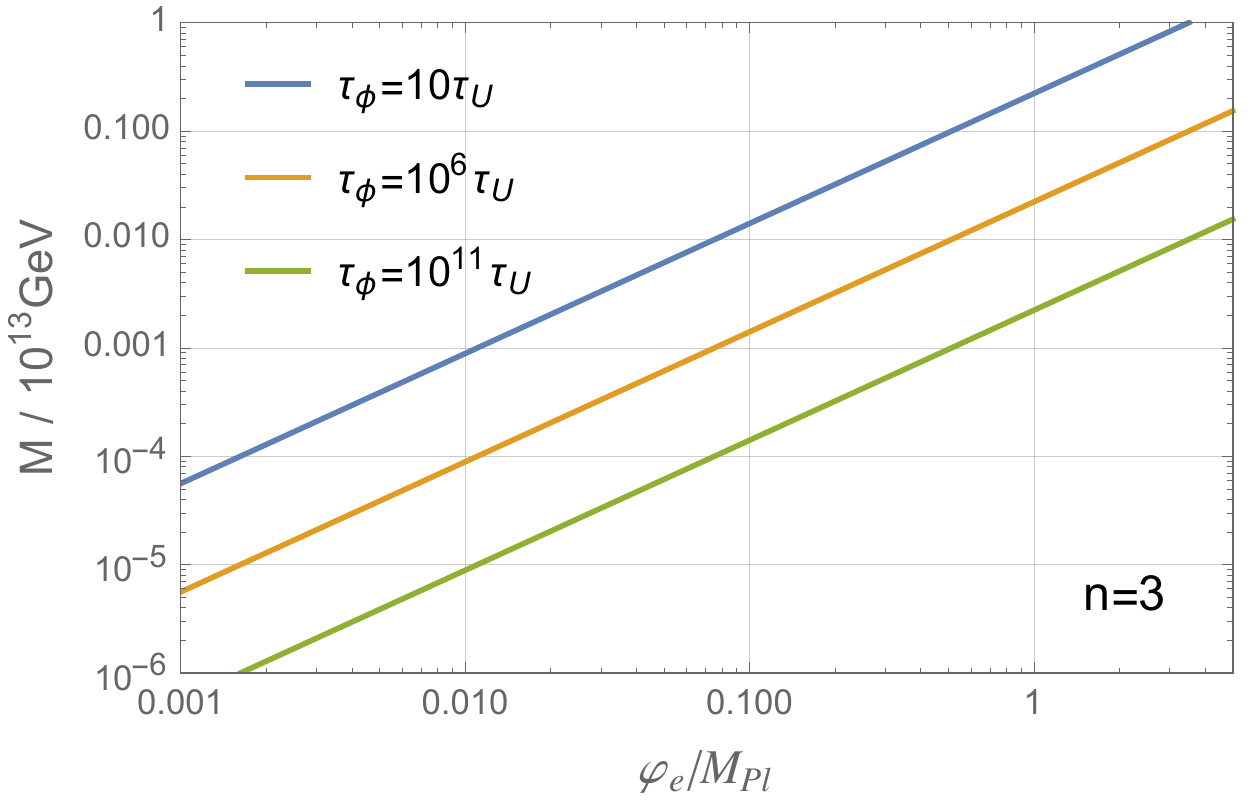}
\includegraphics[width=0.45\columnwidth,angle=0]{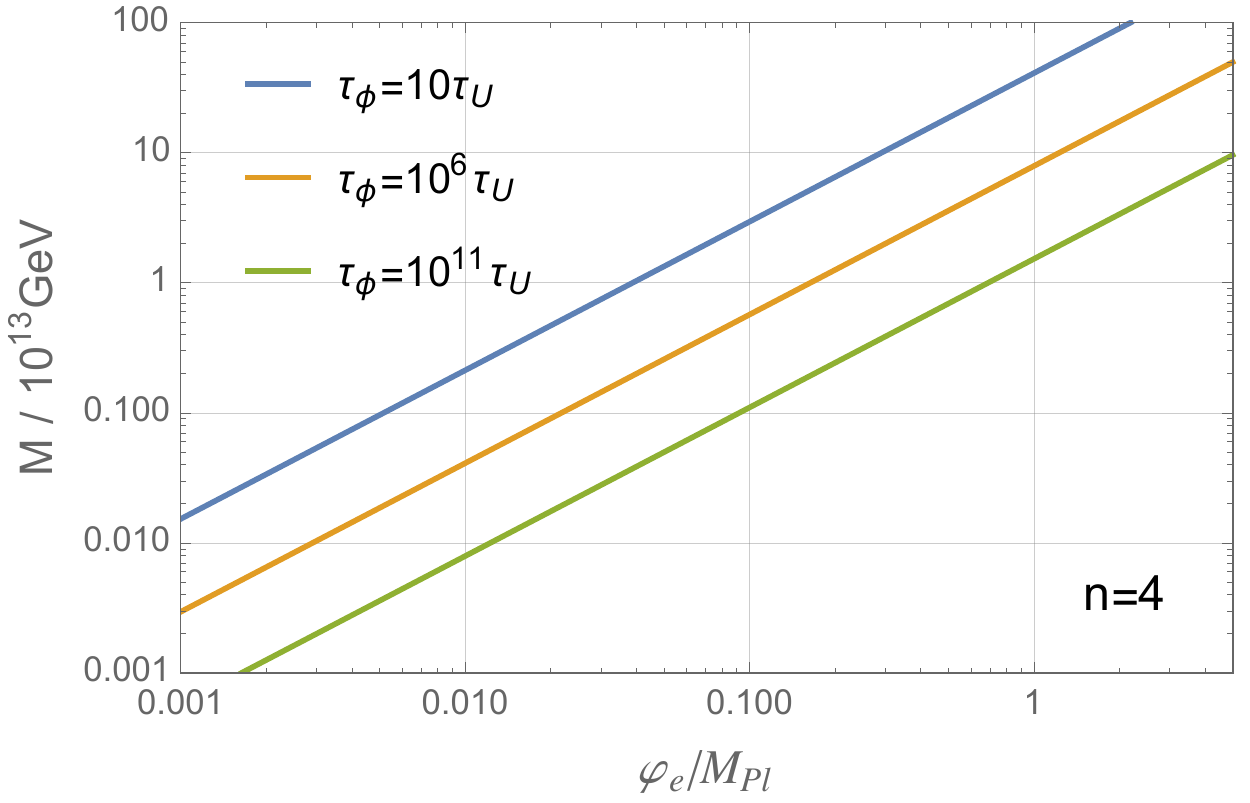}
\includegraphics[width=0.45\columnwidth,angle=0]{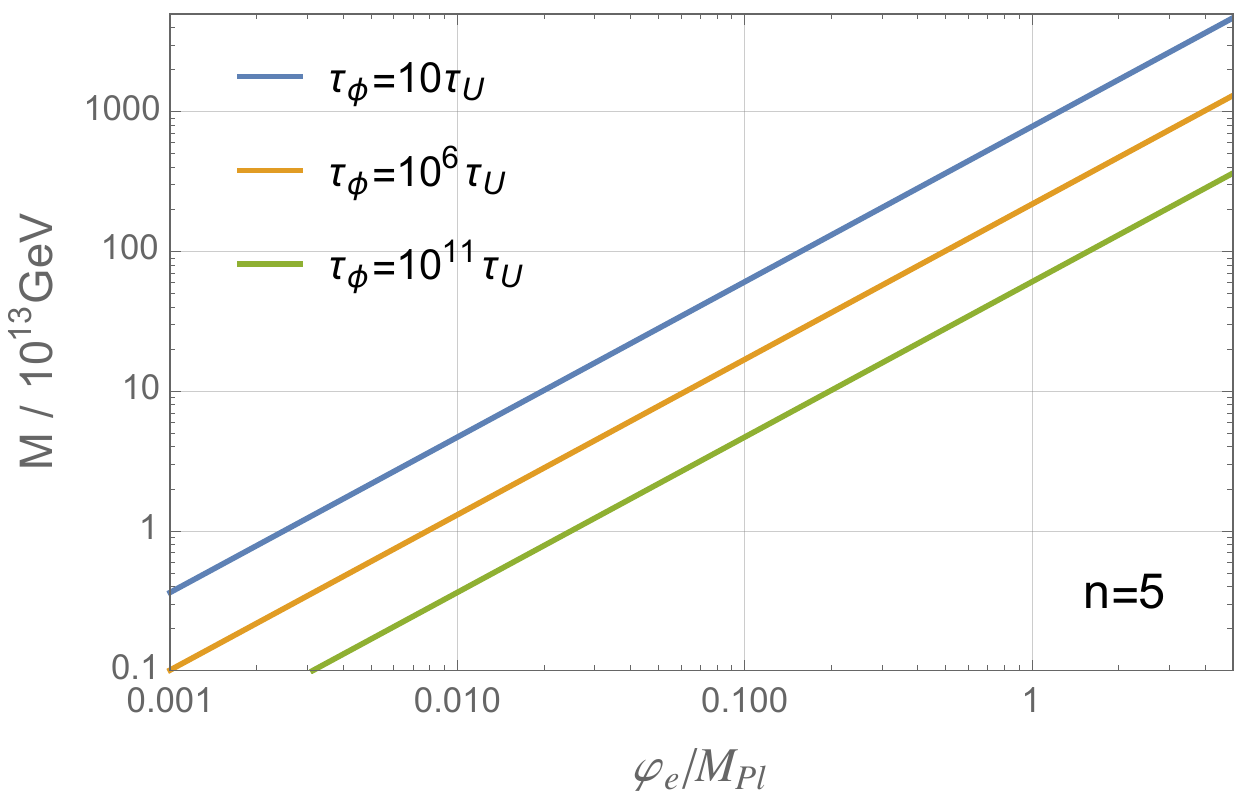}
\includegraphics[width=0.45\columnwidth,angle=0]{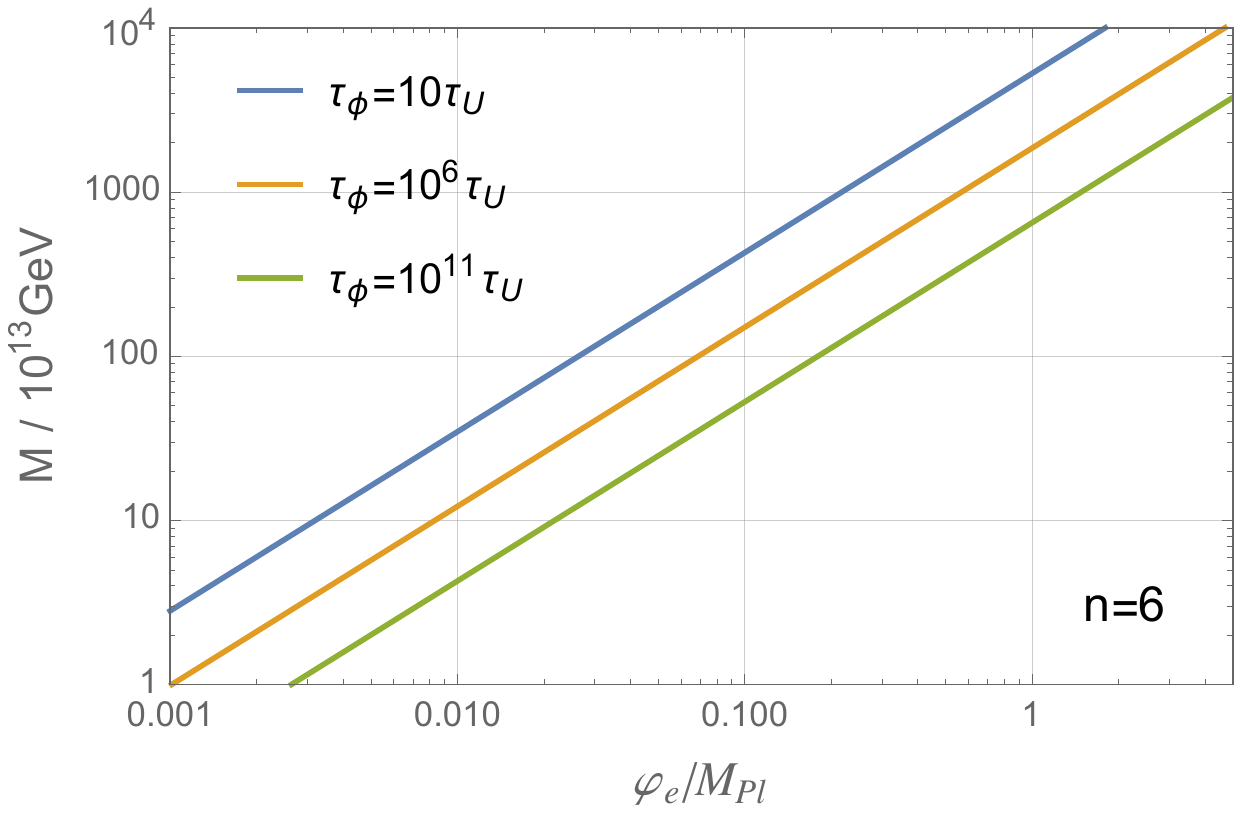}
\caption{Dependence of the scalar $\phi$ mass $M$ on the inflaton field
  $\varphi_e$ at the end of inflation for a set of the DM lifetimes
  $\tau_{\phi}$ and coupling terms \eqref{renorm}. The cases $n=3,4,5,6$ are shown on the top left, top right, bottom left and bottom right plots. We have set $w=0$ (matter-dominated evolution during 
preheating), $A=\frac{H_e}{M}$, and 
$T_{reh} = 10^{15}~\mbox{GeV}$. The region above the blue line
($\tau_{\phi}=10\,\tau_U$) is excluded by the DM stability
constraints. Recall, that we always assume that $H_e\lesssim
  M/5$. Since in a particular model $\varphi_e$ and $H_e$ are related,
some regions on the plots may be unsuitable, if the inequality becomes
invalid.}\label{limits}
\end{center}
\end{figure}

We show the dependence of $M$ on $\varphi_e$ for different fixed values of the DM lifetime $\tau_{\phi}$ in Fig.~\ref{limits}. We set $w=0$, in which case 
the dependence on the Hubble rate $H_e$ drops out according to the estimate~\eqref{noninst}. This is baring in mind the fact that our discussion is applicable only 
for superheavy fields $M >H_e$. In line with the model-independent treatment of inflation, we assume that for any point $(M, \varphi_e)$ one can 
construct the inflationary scenario, which fulfills this inequality.

A couple of comments are in order here. For $n=3$ (renormalizable
 interaction with the inflaton \eqref{renorm}) and $w=0$ one obtains from Eq.~\eqref{rate-0}
\begin{equation}
\nonumber 
\frac{\tau_{\phi}}{\tau_U} \approx 4 \cdot 10^{-6} \cdot \left(\frac{10^{13}~\mbox{GeV}}{M} \right)^5 \cdot \left(\frac{\varphi_e}{M_{Pl}} \right)^{6} \cdot \frac{T_{reh}}{10^{12}~\mbox{GeV}}\; ,
\end{equation}
where we assumed $A=H_e/M$, see Eq.~\eqref{noninst}. For the reference values $\varphi_e = M_{Pl}$, $M = 10^{13}~\mbox{GeV}$,
and $T_{reh} = 10^{15}~\mbox{GeV}$ we get $\tau_{\phi} \approx 0.4 \cdot 10^{-2}
\tau_{U}$, see Fig.~\ref{limits}. Recall that $H_e\lesssim M/5$; then at $\sim 60$ e-foldings before the end of
inflation the Hubble rate $H_{60e}\gtrsim H_e$ is constrained by the absence of the tensor modes,
$H_{60e}\lesssim 8\cdot 10^{13}$\,GeV~\cite{Akrami:2018odb}. It means that this particular
scenario may be just consistent with the DM stability for the Planckian
field inflationary scenarios developing  tensor modes detectable in the future experiments. Thus,
the future detection of primordial gravitational waves would imply
severe limits on this scenario: high
reheating temperature, DM as heavy as $M \simeq 10^{13}~\mbox{GeV}$, and large
field at the end of inflation, $\varphi_e\gtrsim M_{Pl}$, for the
case of the renormalizable interaction, $n=3$. It is straightforward to check
that the same conclusion holds for the radiation-like evolution right
after inflation.

Cosmologically viable part of the model parameter space grows with the dimension of the inflaton coupling function $F(\varphi)$, as it is clearly seen from Fig.~\ref{limits}. 
In particular, for $n=4$, $M=10^{13}~\mbox{GeV}$, $\varphi_e =M_{Pl}$, $A=H_e/M$, and $T_{reh}=10^{15}~\mbox{GeV}$, one gets from Eq.~\eqref{rate-0}: $\tau_{\phi} \approx 
1.5 \cdot 10^{12} \tau_{U}$, which is definitely consistent with the
DM stability. The higher the operator dimension, the longer the DM lifetime in
the model is.

Note that the stability at the time scale of the age of the present
Universe is a necessary but not always a sufficient condition of
viability of the model. DM couples to the inflaton, which in turn 
must couple to the SM particles to reheat the Universe. Hence, even
rarely decaying DM contributes to the cosmic ray flux, which
has been measured in a wide energy range up to $10^{11}$\,GeV. A
reasonable consistency of this flux with expectations from the
astrophysical sources places limits on the decay rate of a heavy relic
of a given mass depending on its decay pattern. Generically, the DM
lifetime must exceed the age of the Universe by many orders of
magnitude. Namely, $\tau_{DM} \gtrsim 10^{20-22}~\mbox{years}$ for $M 
\simeq 10^{13}~\mbox{GeV}$, if the decay initiates a noticeable energy release into gamma
rays or neutrinos, see, e.g., Ref.~\cite{Esmaili:2012us, Kachelriess:2018rty}. This
requirement (if applicable) is consistent with our mechanism for
integer $n>3$ (non-renormalizable interaction \eqref{renorm}), including 
inflationary models predicting potentially observable tensor
modes. The renormalizable model with $n=3$ in Eq.~\eqref{renorm} is
consistent only for small Hubble inflation. So, no detectable relic
gravitational waves are expected in that case.

The value $\varphi_e \simeq M_{Pl}$ is typical in monomial large field
inflation, where the inflaton is minimally coupled to
gravity. Generically, however, the inflaton may substantially deviate
from the Planckian value. An example of this situation is exhibited in the Higgs inflation~\cite{Bezrukov:2007ep,Ema:2017rqn,Gorbunov:2018llf}---one of currently favored models. At the end of inflation, the Higgs field defined in the Jordan frame 
has the value $\varphi_e \simeq \frac{M_{Pl}}{5\sqrt{\xi}}$, where $\xi$ measures the non-minimal 
coupling to gravity; typically $\xi \sim 10^{4}$; we set
$\varphi_e=M_{Pl}/500$. Taking $M =5\cdot 10^{13}~\mbox{GeV}$ and
$T_{reh}=10^{15}~\mbox{GeV}$\,\cite{Bezrukov:2019ylq}, we get from Eq.~\eqref{rate-0}: $\tau_{\phi} \approx  60~\tau_U$, $\tau_{\phi} \approx 10^{10}~\tau_{U}$, and $\tau_{\phi} \approx 2 \cdot 10^{18}~\tau_U $ for $n=6, 7, 8$,
respectively. We see that consistency with cosmic rays propagation requires rather high dimension operators, $n \geq 7$, in the case of Higgs inflation.

So far, we mainly discussed very heavy DM, $M \gtrsim 10^{13}~\mbox{GeV}$, being interested in
inflationary models with detectable relic gravitational waves which
amplitude at production is $\propto H_{e}/M_{Pl}$. However, with the
current null result in searches of primordial gravitational waves, it
is legitimate to consider inflation with a low expansion rate $H$
and masses $M \ll 10^{13}~\mbox{GeV}$. Then any couplings to the inflaton given by Eq.~\eqref{renorm},
including the renormalizable one, $n=3$, become consistent with the
cosmic ray observations. 
In this regard, 
the region $M \lesssim 10^{9}~\mbox{GeV}$ corresponding to the lifetime $\tau_{\phi} \simeq 10^{11}-10^{12} \tau_{U}$ can be of particular interest from the viewpoint of IceCube neutrino observations~\cite{Feldstein:2013kka, Esmaili:2014rma}. 
Namely, if the inflaton mainly decays into leptons, one can explain the origin of PeV neutrinos without spoiling Fermi limits on the gamma rays~\cite{Kachelriess:2018rty} obtained in Refs.~\cite{Cohen:2016uyg, Blanco:2018esa}. Note that the region of interest corresponds to relatively small values of the inflaton at the end of inflation: $\varphi_e \lesssim 0.1 M_{Pl}$ in the renormalizable 
case $n=3$. Even smaller values of $\varphi_e$ are required for $n>3$.

We finish this Section with two concluding remarks. First, let us estimate the typical value of the coupling constant $\alpha$. Taking $M = 10^{13}~\mbox{GeV}$, $\varphi_e = M_{Pl}$, and $T_{reh} = 10^{15}~\mbox{GeV}$ and 
assuming $\Lambda =M_{Pl}$ in Eq.~\eqref{renorm}, we get from Eq.~\eqref{abundance0} 
\begin{equation}
\label{est}
\alpha \approx 0. 3 \cdot 10^{-24} \; .
\end{equation}
Thus, our mechanism implies extremely feeble interactions with the
inflation. 

Second, for fixed $\tau_{\phi}/\tau_{U}$ and $\varphi_e$, 
one can estimate the maximal possible value of $M$ achieved in the
limit $n \rightarrow \infty$ (both Eqs.~\eqref{rate-0} and \eqref{rate-1/3}
give the same result):
\begin{equation}
\nonumber 
\frac{M}{10^{13}~\mbox{GeV}} \simeq 10^{7} \; \sqrt{n} \cdot \frac{\varphi_e}{M_{Pl}} \; .
\end{equation}
Hence, for the large field inflation, there is essentially no upper
bound on the mass of the field $\phi$ produced. Say, $M=M_{Pl}$, the
largest mass allowed within the quantum field theory at our present
understanding of gravity, is achieved with $n \approx 18$ for $\varphi_e=M_{Pl}$ and $\tau_{\phi}\sim10^{12} \cdot \tau_{U}$. This implies that the decaying DM
can contribute to the cosmic rays starting from the Planckian
energies, that may be observed (at least in the neutrino sector, where
the energy does not degrade).

\section{Scenario with subsequent decay to lighter particles}

In the rest of the paper, we discuss a variation of our basic scenario assuming that the superheavy fields 
are unstable, while the ''true'' DM particles appear as their decay
products. This is the only viable option in the inflationary scenarios
with inherently short lifetime  
of particles $\phi$.

 \begin{figure}[tb!]
\begin{center}
\includegraphics[width=0.45\columnwidth,angle=0]{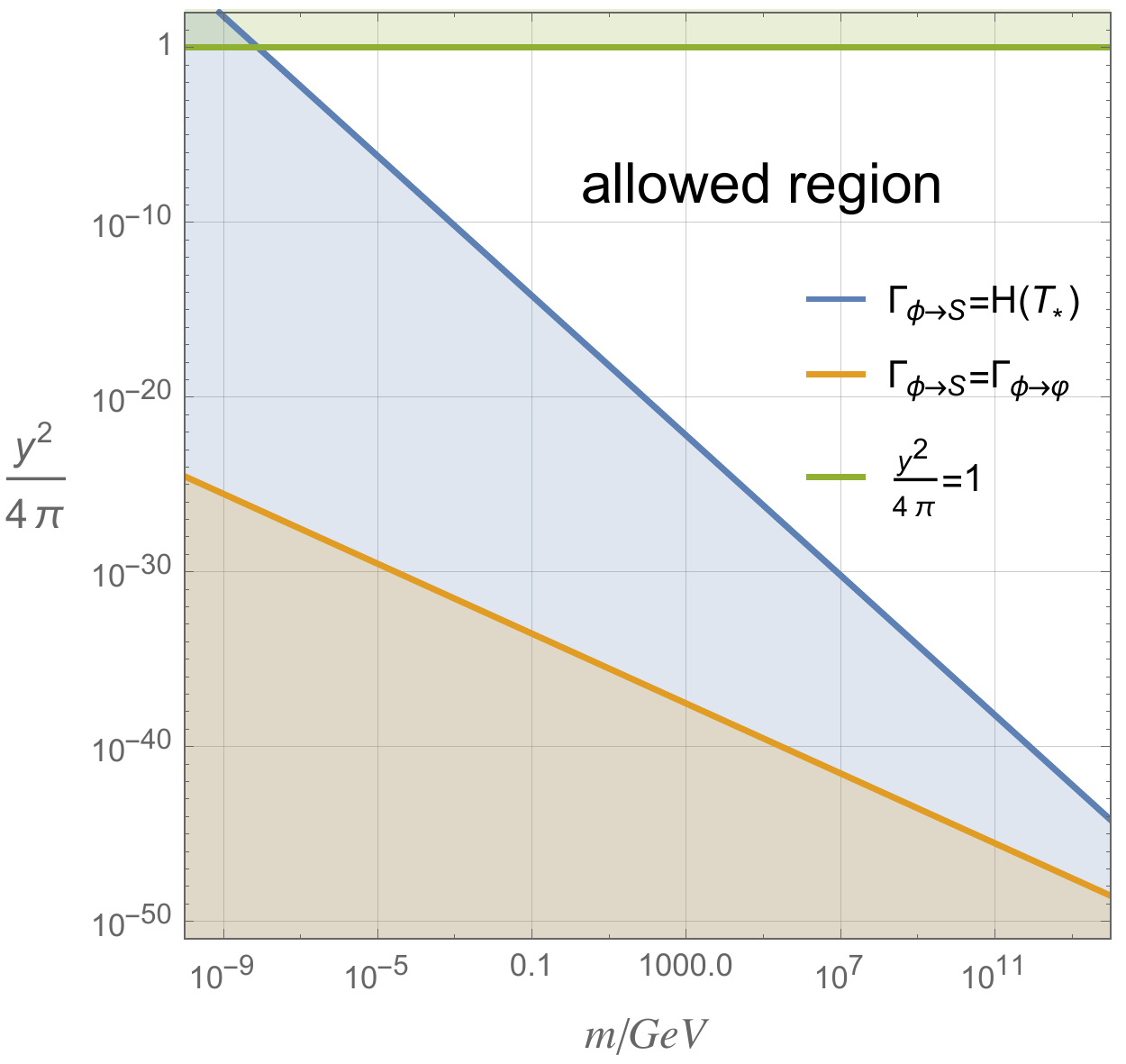}
\includegraphics[width=0.45\columnwidth,angle=0]{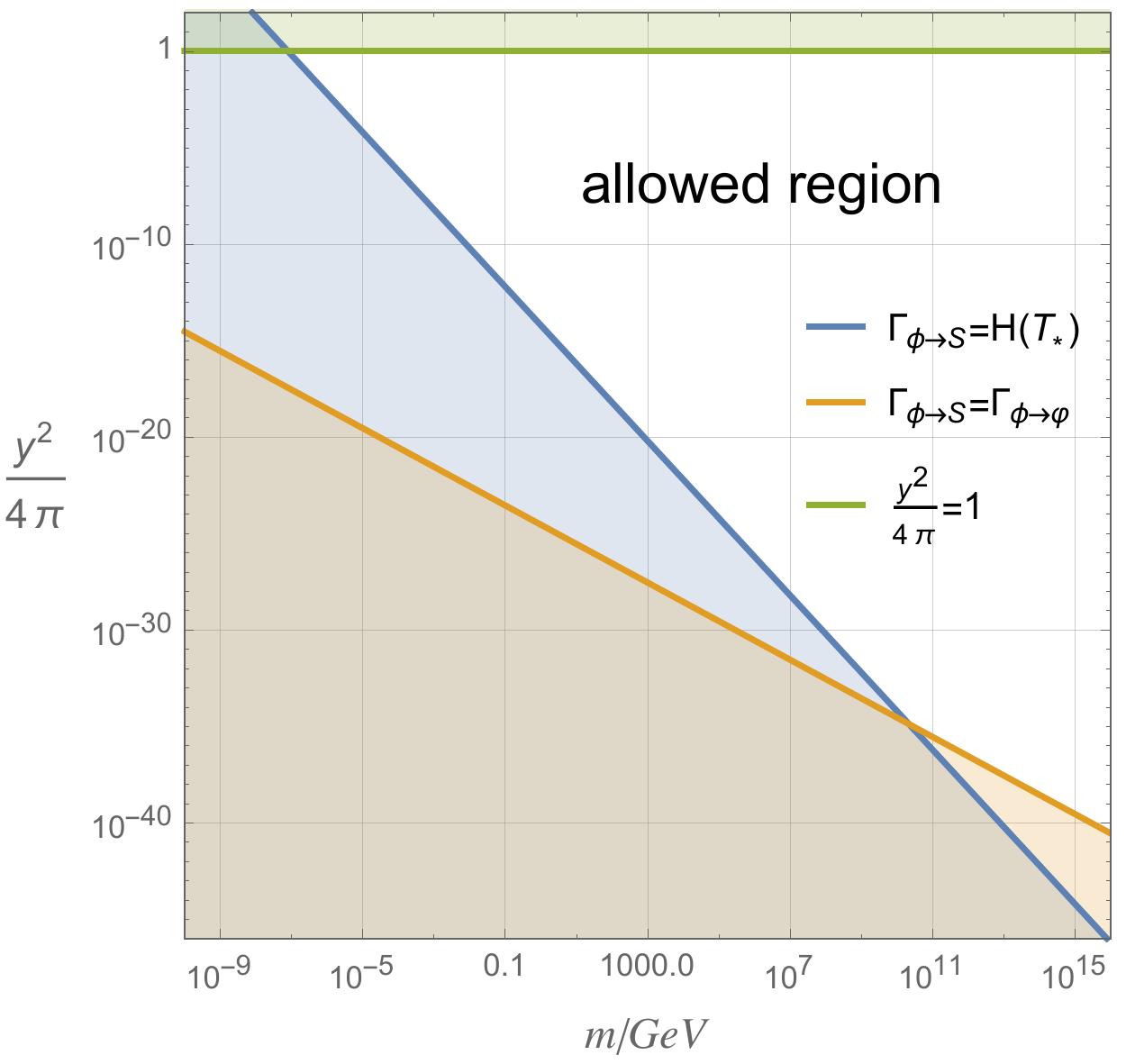}
\caption{The plots show allowed values of the coupling constant $\frac{y^2}{4\pi}$ and the DM mass $m$ in the scenario, where the superheavy field $\phi$ produced 
during inflation subsequently decays to a couple of stable Dirac
fermions. We have set $w=0$ (matter-dominated evolution right after
inflation), $\varphi_e=M_{Pl}$,  $T_{reh} =10^{15}~\mbox{GeV}$,
$A=H_e/M$. For the field $\phi$ mass we put $M=10^{14}~\mbox{GeV}$
(left plot) and $M=10^{16}$~\mbox{GeV} (right plot). The light blue
regions 
correspond to cosmologically unacceptable hot DM.}\label{limitslight}
\end{center}
\end{figure}

We assume that the field $\phi$ has an additional Yukawa coupling to
the Dirac fermion $S$ of the mass $m$ --- 
a singlet with respect to the SM gauge group, 
\begin{equation}
\label{light}
{\cal L}=y \phi \bar{S} S \,, 
\end{equation}
where $y$ is a dimensionless Yukawa coupling. The Dirac fermion is stable
and serves as DM. In this picture,
the concentration of DM particles is still fixed by the inflationary
dynamics, and it is twice that of the particle $\phi$:  
\begin{equation}
\label{abundance1}
 n_{DM} = \frac{A^2 F^2(\varphi_e)}{M^3} \cdot \left( \frac{a_e}{a}\right)^3 \; .
\end{equation}
The energy density of DM is then given by $\rho_{DM} =m \cdot
n_{DM}$. The condition that it 
constitutes (almost) all of the invisible matter in the Universe reads
\begin{equation}
\label{abundance0light}
\frac{8\pi \, h_{*} (T_{eq})\, A^2 \, F^2_e \,
  T_{reh}}{3 \, g_* (T_{eq})\, M^2 \, H^2_e \, M^2_{Pl}
  \, T_{eq}}\cdot \frac{m}{M} \approx 1 \; ,
\end{equation}
where we have chosen the scenario with the matter dominated evolution
right after inflation, cf. Eq.~\eqref{abundance0}. Note that in this version of the mechanism the coupling constant 
of the scalar $\phi$ to the inflaton can be substantially larger
compared to the estimate~\eqref{est} by the factor $\sqrt{M/m}$.

The particles $S$ produced in the decays of the scalar $\phi$ generically have very high momenta $\simeq M/2$ at the moment of decay. 
On the other hand, DM particles must be very non-relativistic at the matter-radiation equality: the velocity of DM fluid should not exceed $v \simeq 10^{-3}$. 
Otherwise, a well established picture of the large scale structure formation would be spoiled. 
In order to fulfill this condition, the particles $S$ must become non-relativistic at least by the time, when the Universe cools down to $T \simeq 1~\mbox{keV}$. Hence, the scalar $\phi$ should decay into 
the particles $S$ before the Universe temperature reaches
\begin{equation}
\nonumber 
T_{*}=\l\frac{M}{2m}\r \times 1\; \text{keV} \; .
\end{equation}
That is, the following condition must be obeyed: 
\begin{equation} 
\label{lightconstr1}
\Gamma_{\phi \rightarrow S} \gg H(T_*) =\sqrt{\frac{8\pi^3 g_*(T_{*})}{90}} \cdot \frac{T^2_{*}}{M_{Pl}} \; .
\end{equation}
Furthermore, the decay rate into $S$-particles must exceed that into the inflatons, i.e., 
\begin{equation}
\label{lightconstr2}
\Gamma_{\phi \rightarrow \varphi} \ll \Gamma_{\phi \rightarrow S} \; .
\end{equation}
If there is the decay in two light particles, as it is suggested by Eq.~\eqref{light}, then its rate is given by 
\[
\Gamma_{\phi \rightarrow S} =\frac{y^2M}{8\pi}\,.
\]
The decay rate $\Gamma_{\phi \rightarrow \varphi}$ is inferred from Eq.~\eqref{nonrenorm}. We assume the renormalizable interaction with the inflaton, i.e., $n=3$. 
The conditions Eqs.~\eqref{lightconstr1} and~\eqref{lightconstr2} can be interpreted as the constraints on the coupling constant $y$: 
\begin{equation}
\label{inequality1}
2 \cdot 10^{-26} \cdot \sqrt{\frac{g_* (T_{*})}{10}} \cdot \left(\frac{M}{10^{13}~\mbox{GeV}} \right) \cdot \left(\frac{10~\mbox{TeV}}{m} \right)^2 \ll \frac{y^2}{4\pi} \ll 1 \; ,
\end{equation}
and 
\begin{equation}
\label{inequality2}
\frac{y^2}{4\pi} \gg 3 \cdot 10^{-41}  \cdot \left(\frac{H_e}{M \cdot A} \right)^2 \cdot \left(\frac{M}{10^{13}~\mbox{GeV}} \right)^5 \cdot \left(\frac{M_{Pl}}{\varphi_e} \right)^6 \cdot \left( \frac{10^{12}~\mbox{GeV}}{T_{reh}} \right) \cdot \frac{10~\mbox{TeV}}{m} \; .
\end{equation}
None of the above constraints is particularly restrictive leaving a
broad range of possible values of the coupling constant $y$ for fairly arbitrary masses $m$, as is shown in Fig.~\ref{limitslight}. 
Interestingly, when the value of $y^2/4\pi$ is close to its lower
bound in the inequality~\eqref{inequality1}, warm DM is produced. 
This is despite the fact that the DM particles can be heavy, well 
above $\sim \mbox{keV}$, cf. Ref.~\cite{Gorbunov:2008ka}. Note that we again assume the estimate~\eqref{noninst} for the 
coefficient $A$. More generic powerlaw dependence of the coefficient $A$ is not expected to change the 
picture dramatically. However, if $A$ decreases exponentially with $M$, the present analysis 
should be revisited.

\section{Discussions}

We studied in detail a novel mechanism of producing superheavy DM in the form of the scalar 
field $\phi$ condensate. For 
any given inflationary model and the coupling of the field $\phi$ to the inflaton, 
the DM decay rate can be calculated, and the results can be contrasted with the existing data on the propagation of the cosmic rays. 
The choice of dataset one should use depends on the composition of cosmic rays originating from the 
decays of the inflaton. 
In turn, the composition depends on the interaction of the inflaton
with the SM particles, responsible for the reheating in the early Universe.

For the simplest possible couplings of the field $\phi$ to the
inflaton, our mechanism is 
very predictive, allowing to exclude a set of inflationary scenarios (provided that the mechanism works), 
or strongly constrain the range of DM masses. For example, the renormalizable interaction is only marginally consistent with the DM stability constraint, $\tau_{DM} \gtrsim 10^{10}~\mbox{years}$, 
in the high scale inflationary scenarios with the Hubble rate $H \simeq 10^{13-14}~\mbox{GeV}$. Hence, 
possible future observation of gravitational waves will strongly corner this option. The parameter space is broader in the case of 
non-renormalizable interactions. 

On the other hand, if searches for tensor modes show null result, the window for possible masses $M$ is essentially unbounded from below. 
The typical DM lifetime can be very large in that case. If $\tau_{\phi} \gtrsim 10^{20-22}~\mbox{years}$, one can entertain the 
opportunity that a fraction of the observed very high energy neutrinos and gamma-rays originate from the decays of DM.

More generally, the scenario considered in the present work, can be viewed as a mechanism of generating 
superheavy fields---not necessarily DM. Subsequent decays of these fields may source DM in the form of 
some lighter stable particles from the Standard Model extensions, e.g., sterile neutrinos. Alternatively, 
these fields can be used for creating baryon asymmetry in the Affleck--Dine fashion~\cite{Babichev:2018sia}.  
This opens up the opportunity of unified description of DM production and baryogenesis.

\section*{Acknowledgments} We are indebted to Mikhail Kuznetsov, Lorenzo Reverberi, and Federico Urban for useful discussions. 
E.B. acknowledges support from the research program ``Programme national de cosmologie et galaxies'' of the CNRS/INSU, France, 
and from 
Projet de Recherche Conjoint n\textsuperscript{o}~PRC 1985: ``Gravit\'e modifi\'ee et trous noirs: signatures exp\'erimentales et mod\`eles consistants'' 2018--2020.
S.R. is supported by the European Regional Development Fund (ESIF/ERDF) 
and the Czech Ministry of Education, Youth and Sports (M\v SMT) through the Project CoGraDS-CZ.02.1.01/0.0/0.0/15\_003/0000437.

\section*{Appendix} 

In the present Appendix, our goal is to estimate the coefficient $A$ entering Eq.~\eqref{condensate} using different toy examples. We simplify the problem by switching off the Hubble friction in Eq.~\eqref{osc}. Then, the equation, which describes the evolution of the 
field $\phi$ is given by  
\begin{equation}
\nonumber 
\ddot{\phi}+M^2 \phi=F(t) \; .
\end{equation}
The Hubble friction serves to set the tracking solution for the field 
$\phi$ during inflation. However, the same can be achieved by tuning initial conditions: 
\begin{equation}
\label{incond}
\phi=\frac{F(\varphi)}{M^2} \qquad \dot{\phi}=\frac{\dot{F} (\varphi)}{M^2} \qquad (t \rightarrow -\infty) \; .
\end{equation}
It is convenient to split the solution for $\phi$ in two parts:  
\begin{equation}
\nonumber 
\phi=\frac{F}{M^2}+\chi \; ,
\end{equation}
The first term on the r.h.s. decays together with the inflaton. We are interested in the second term, $\chi$, which is relevant for the observed abundance 
of DM in our picture. The field $\chi$ satisfies the equation:  
\begin{equation}
\nonumber 
\ddot{\chi}+M^2 \chi=-\frac{\ddot{F}}{M^2} \; .
\end{equation}
This equation is to be solved with the trivial initial conditions: $\chi (t \rightarrow -\infty)=0$ and $\dot{\chi} (t \rightarrow -\infty)=0$, which match~\eqref{incond}. 
The result reads: 
\begin{equation}
\label{chi}
\chi=\frac{\cos (Mt)}{M^3} \cdot \int^{t}_{-\infty} dt' \sin (Mt') \ddot{F}(t')-\frac{\sin (Mt)}{M^3} \cdot \int^{t}_{-\infty} dt' \cdot \cos (Mt') \ddot{F}(t') \; .
\end{equation}
We are interested in large times $t$, when $F(\varphi) \rightarrow 0$ and consequently $\phi \rightarrow \chi$. In this limit, one can also replace the limits of integration by $\infty$. One gets
\begin{equation}
\label{remnant}
\phi (t)=\frac{\cos (Mt)}{M^3} \cdot \int^{+\infty}_{-\infty} dt' \sin (Mt') \ddot{F}(t')-\frac{\sin (Mt)}{M^3} \cdot \int^{+\infty}_{-\infty} dt' \cdot \cos (Mt') \ddot{F}(t') \; .
\end{equation}
This is the field value, which feeds into the observed abundance of DM. We see that it is never zero generically. The value of the coefficient $A$ can be read off from Eq.~\eqref{remnant} by the comparison with Eq.~\eqref{condensate}. 
We see that $A$ is defined by the Fourier transform 
of the function $\ddot{F}$. We have no a priori expectations about the $M$-dependence of the Fourier transform. However, the examples below 
point out that the coefficient $A$ i) drops exponentially with $M$ for
a smooth slow evolution of the inflaton 
including its derivatives; ii) has a powerlaw dependence on $M$, if some derivatives of the inflaton change very fast on the time scale $M^{-1}$.

{\it Exponential behavior of $A$.} To illustrate the former situation, let us choose the function $F$ as follows: 
\begin{equation}
\nonumber 
F(t)=\Lambda^3 \left[\frac{1}{2}-\frac{1}{\pi} \arctan (m t) \right]\; ,
\end{equation}
where $\Lambda$ and $m$ are two parameters of the mass dimension. In the limit $t\rightarrow -\infty$ one has $F(t \rightarrow -\infty) =\Lambda^3=\mbox{const}$, and at $t \rightarrow +\infty$ one has $F(t) \rightarrow 0$. 
Hence, this function $F$ correctly captures dynamics of the inflaton, which is nearly constant initially and then decays at the post-inflationary epoch. The second derivative 
of the function $F$ is given by 
\begin{equation}
\nonumber 
\ddot{F}=\frac{2m^3 \Lambda^3 t}{\pi \left[(1+(mt)^2 \right]^2} \; .
\end{equation}
One calculates the integrals in Eq.~\eqref{chi} using Jordan's lemma. The result reads
\begin{equation}
\nonumber 
\phi (t)=\frac{\Lambda ^3 \cdot \cos (Mt)}{M^2} \cdot e^{-\frac{M}{m}}  \; .
\end{equation}
Comparing with Eq.~\eqref{condensate}, we obtain finally 
\begin{equation}
\nonumber 
A=2 \cdot e^{-\frac{M}{m}} \; .
\end{equation}
Hence, the coefficient is of order one for an abrupt change, $m\gtrsim M$
or exponentially small for slow evolution, $m\ll M$.

{\it Powerlaw behavior of $A$.} Now let us consider the situation, when the $n$-th derivative of the function $F(t)$ changes abruptly, so that 
$F^{(n+1)} \propto \delta (t)$. In practice, it is enough, if $F^{(n+1)}$ changes fast on the times scales $M^{-1}$. In this 
situation, by integrating Eq.~\eqref{remnant} by parts, we get
\begin{equation}
\nonumber 
\phi (t)=\frac{\cos (Mt +\delta)}{M^{n+2}} \cdot F^{(n)} (0) \; ,
\end{equation}
where $\delta$ is an irrelevant phase. Comparing with Eq.~\eqref{condensate}, we read off the coefficient $A$: 
\begin{equation}
\label{estim}
A \simeq \frac{F^{(n)} (0)}{M^n \cdot F (0)} \; .
\end{equation}

Now let us support the estimate~\eqref{estim} using an example, which incorporates the effects 
of the Hubble friction. It is easy to model the situation, where the first derivative of the inflaton 
undergoes a discontinuous jump. For this purpose, we choose the inflaton profile as follows  
\begin{equation}
\varphi=\varphi_e =\mbox{const}~(t<t_e) \qquad \varphi =\varphi_e \cdot \left(\frac{a_e}{a} \right)^{3/2} \;  .
\end{equation}
Namely, inflation is approximated by the exact de Sitter stage followed by the post-inflationary epoch with the equation of state $w=0$ and the scale 
factor $a(t) \propto t^{2/3}$. During the matter-dominated stage, the exact general solution for the field $\phi (t)$ can be written as 
\begin{equation}
\nonumber 
\phi(t)=-\frac{\cos (Mt)}{a^{3/2} (t)} \cdot \int dt \frac{\sin (Mt)}{M} F(\varphi) a^{3/2} (t)+\frac{\sin (Mt)}{a^{3/2}} \cdot \int dt \frac{\cos (Mt)}{M} F(\varphi) a^{3/2} (t) \; .
\end{equation}
Up until the times $t=t_e$, the field $\phi$ tracks the inflaton, so that its initial conditions at the onset of the post-inflationary decay are given by  
\begin{equation}
\label{initial}
\phi(t_e)=\frac{F(\varphi_e)}{M^2} \qquad \dot{\phi} (t_e)=0 \; .
\end{equation}
As in the bulk of the paper, we assume that $F(\varphi)$ is a powerlaw, i.e., $F(\varphi) \propto \varphi^n$. Now we are ready to write down the solution, which satisfies the initial conditions~\eqref{initial}:
\begin{equation}
\nonumber 
\begin{split}
\phi &=\frac{F_e \cos \xi}{M^2} \cdot \frac{Mt_e}{Mt_e+\xi} \cdot \left[1-\int^{\xi}_0 d\xi' \frac{(Mt_e)^{n-1} \sin \xi'}{(Mt_e+\xi')^{n-1}} \right]+ \\
&+\frac{ F_e \sin \xi}{M^2} 
\cdot \frac{Mt_e}{Mt_e+\xi} \cdot \left[\int^{\xi}_0 d\xi' \cdot \frac{(Mt_e)^{n-1} \cos \xi'}{(Mt_e+\xi')^{n-1}}+\frac{1}{Mt_e} \right] \; ,
\end{split}
\end{equation}
where $\xi \equiv M(t-t_e)$. For the sake of concreteness, we focus on the case $F(\varphi) \propto \varphi^2$. We use the fact that the integrals over $\xi$ are converging fast, and thus can be replaced by their values at large $\xi$:
\begin{equation} 
\int^{\infty}_0 d\xi' \cdot \frac{Mt_e \cos \xi'}{Mt_e+\xi'}=\frac{1}{Mt_e} \qquad \int^{\infty}_0 d\xi' \cdot \frac{Mt_e \sin \xi'}{Mt_e+\xi'}=1+{\cal O} \left(\frac{1}{[Mt_e]^2} \right)  \; .
\end{equation}
Hence, the solution of interest reads 
\begin{equation}
\phi (t \gg t_e)=\frac{2F_e}{M^3t_e} \cdot \left(\frac{a_e}{a} \right)^{3/2} \cdot \sin [M(t-t_e)] \; . 
\end{equation}
Comparing the latter with Eq.~\eqref{condensate}, we get for the coefficient $A$: 
\begin{equation}
\nonumber 
A =\frac{2}{Mt_e} \simeq \frac{H_e}{M} \; .
\end{equation}
This is a cross-check of Eq.~\eqref{estim} and justification of Eq.~\eqref{noninst}. We have checked that this result is robust against different 
choices of the function $F(\varphi)$ (still, powerlaw in $\varphi$). Furthermore, we have also considered the situation, when the 2-nd derivative of the 
inflaton experiences a discontinuous jump, while the $0$-th and the $1$-st ones are smooth. In that case, one gets $A \simeq H^2_e/M^2$ in agreement with Eq.~\eqref{estim}.


\end{document}